# Human pitch is pre-cortical: The essential role of the cochlear fluid

Florian Gomez and Ruedi Stoop[1]

[1]Institute of Neuroinformatics, University and ETH Zurich, 8057 Zurich, Switzerland

(Dated: November 5, 2013)

## Abstract

The perceived pitch of a complex harmonic sound changes if the partials of the sound are frequency-shifted by a fixed amount. Simple mathematical rules that the perceived pitch could be expected to follow ('first pitch-shift') are violated in psychoacoustic experiments ('second pitchshift'). For this, commonly cognitive cortical processes were held responsible. Here, we show that human pitch perception can be reproduced from a minimal, purely biophysical, model of the cochlea, by fully recovering the psychoacoustical pitch-shift data of G.F. Smoorenburg (1970) and related physiological measurements from the cat cochlear nucleus. For this to happen, the cochlear fluid plays a distinguished role.

Pitch is a central and most intriguing trait of human hearing. Since the phenomenon of combination tones was discovered and used in musical compositions (Sorge 1745, Tartini 1754), understanding pitch on a physical basis has become an obsession to great physicists, such as Helmholtz, Ohm and Seebeck. The aim of this letter is to demonstrate that pitch is indeed an entirely physical - as opposed to cortical - phenomenon. Previously, we have demonstrated that if the cochlear signal is described in terms of local wave-forms, the *ingredients* of the perceived pitch are present at the cochlear level. The precise physical basis responsible for this, and how the perceived pitch should be extracted, remained, however, unexplained. Here we reveal the precise physical pitch-generating mechanisms and design a simple device that reproduces pitch as measured by Smoorenburg's psychophysical experiments [1]. Our results open the perspective that many of the unexplained psychophysical phenomena to date may not be cortical, but may be understandable from a careful physical analysis of the necessarily nonlinear nature of the involved sensors.

For *pure tones*, pitch coincides with the physical frequency of the sound. This changes if a tone contains several partial sounds. Although quite successful models of pitch perception have been developed (cf. [2] for a review), the mechanisms of pitch perception of complex sounds are still under dispute. Given a *complex sound* containing N *subsequent harmonics*

$$k f_0, (k+1)f_0, (k+2)f_0, ..., (k+N-1)f_0 \qquad (1)$$

of some fundamental frequency $f_0$ (*i.e.* k > 1), for k not too high and if N ≥ 2, the perceived pitch $f_p$ is the fundamental $f_0$. This *'residue pitch'* or *'missing fundamental'* phenomenon was already known to Seebeck [3]; in the case of N = 2 or N = 3, the residue frequency
coincides with the *modulation frequency* of the signal. For psychoacoustic experiments with more complex sounds, this interpretation, however, fails: if all partials are shifted by a fixed amount $\delta f$ (keeping the modulation frequency $f_0$ fixed), a *shift* of the perceived pitch $f_p$ is observed [4–6] (see Fig. 4a), black stars). Simple phenomenological models [4–6] propose for the pitch shift the rule

$$f_p = f_0 + \frac{\delta f}{k'}, \qquad (2)$$

where for smaller $N$, $k'$ is a loosely defined 'center' of the forcings $\{k, k+1, k+2,...\}$ (for $N = 3$: $k' = k+1$, or, for N = 2, cf. black lines in Fig. 4). For larger $N$, $k'$ is chosen as one of the lower frequencies present. When neuronal threshold oscillators were stimulated by a signal $A(\sin f_1 t + \sin f_2 t + ... + \sin f_N t) + \xi(t)$ with frequency components chosen as in Eq. (1) and Gaussian white noise $\xi(t)$, interspike distributions centered at frequencies $f_p$ as in Eq. (2) were found, with $k' = k + (N - 1)/2$ [7], confirming the idea that the main resonance should be the dominant periodicity of the subsequent maxima in the stimulus waveform. This view parallels the temporal pitch perception paradigm [1, 2], where $f_p$ is inferred from the waveforms' most prominent peak in the *autocorrelation function*, or for the auditory nerve from the peak in the *interspike interval histogram* [8, 9]. For two-tone stimuli ($N$ = 2), a pitch-shift of $\delta f/(k+1/2)$ should emerge. Smoorenburg's two-tone pitchshift experiments (and corresponding measurements from the cat cochlear nucleus) showed, however, that for general complex sounds also this rule fails (cf. Fig. 4a), b), black stars vs. black lines). As the origin of the shift, combination tones (CT, also known as 'distortion products') were suspected [1, 10].

CT are a consequence of cochlear nonlinearity. Given a sequence of harmonics $kf_0, ..., (k + N - 1)f_0$, a cubic nonlinearity as in our cochlea re-introduces all the missing partials (or CT)

$$(k - 1)f_0, (k - 2)f_0,.. \text{ and } (k + N)f_0, (k + N + 1)f_0,.., \qquad (3)$$

where the partials above the stimulus frequencies (c.f. Fig. 1b), c)) are generally not perceivable [11]. The general idea now is that lower-frequency CT may shift the "center of gravity" of the stimulus towards lower frequencies and that this leads to a substantial increase of the slope of the lines $f_p(\delta f)$ in Eq. (2) and thereby to the pitch as perceived by a human listener (cf. Fig. 1a) for the increasing importance of CT along the cochlea). To reveal the physical laws underlying perceived pitch, we will scrutinize in what follows on the behavior of CT, by comparing biological CT to CT generated from a biophysically realistic model of the cochlea. We will show that the model fully complies with the presently accepted pitch-related biophysical data [12]. The link to human-perceived pitch and pitch from the cat cochlear nucleus is ensured by the faithfulness of the auditory nerve in relaying the cochlear perceived pitch [13, 14]. The hearing model then enables us to further investigate what physics principles are involved in the generation of pitch. We will find that the presence of the fluid in the cochlea (that in many cochlea and pitch perception models is ignored), is pivotal for obtaining the pitch as perceived by humans.

*Cochlea model* A decade ago it was suggested [15, 16] that Hopf relaxation oscillators close to bifurcation could account for all salient nonlinear properties of hearing. Based on these suggestions, we designed a biophysically detailed theoretical Hopf cochlea [17, 18], translated this concept into analogue hardware [19], and put the latter into dedicated software [13]. The hardware as well as the software versions of this cochlea reproduce the known biological data extremely well (e.g. [20], Supplemental Materials, [13]). Our cochlea is discretized into sections. Each one is hosting an amplification process that is the result of a stimulated Hopf process

$$dz/dt = (\mu + i)\omega_{ch}z - \omega_{ch}|z|^2z - \omega_{ch}F(t), z \in \mathbb{C}, \qquad (4)$$

followed by a low-pass filter mimicking the frequency-specific viscous losses within the cochlear fluid [17] ($F(t)$ is the stimulation signal and $\mu$ the distance to the Hopf bifurcation point). For all results presented here, we used a software model containing 20 sections with characteristic frequencies from 14.08 to 0.44 kHz (5 octaves). For the first five sections, the bifurcation parameters $\mu$ were set to $-0.1$, afterwards they decrease with $-0.025$/section. This leads to amplification and tuning curves that are congruent with biology also at the more apical sections and provides the optimal basis for the simple pitch extraction method that we will finally demonstrate. Apart from the discretization into sections, the precise correspondence between our Hopf cochlea and biology is established by mapping the respective frequency ranges onto each other. Amplitudes are matched so that -114 dB (Hopf cochlea) corresponds to 0 dB SPL (biology).

*CT-measurements and analytics* CT were originally assumed to be relevant for high sound levels only, as at low to moderate sound levels, the hearing system would be essentially linear (hence their alias 'distortion products'). This view was refuted by psychoacoustic evidence [1, 11, 21], demonstrating that CT are already perceived at relatively low sound levels, and that they are ubiquitously present in the hearing system. Fig. 1a) exhibits the importance of CT in cochlear response. As the result of the cochlea's nonlinearity, generated CT amplitudes exhibit an exponential decay in combination 'order', where the decay exponent also depends on the stimulation amplitude (cf. Fig. 1b), c)). Exponential decaying CT levels were first observed in psychoacoustic experiments [11]. Direct experimental observation of inner ear basilar membranes CT became possible by laser interferometry (Fig. 1 b), [12]). The phenomenological pitch formula's problem with complex sounds is that it inherently assumes partials to contribute with equal amplitudes. CT tones do not meet this assumption.

To understand the exponential decay law of CT, we consider a signal composed of the harmonics of given angular frequencies $k\omega_0, ..., (k+N-1)\omega_0$ and amplitudes $F_k, ..., F_{k+N-1}$ (some amplitudes possibly having a zero value). All CT are multiples of $\omega_0$; the response of a single Hopf-oscillator therefore assumes the Fourier series form $z(t) = \sum_j a_j e^{ij\omega_0 t}$. For a frequency $\omega_l = l\omega_0$, we obtain

$$(i(\omega_l - \omega_{ch}) - \mu\omega_{ch})a_l + c.i.t. = -\omega_{ch}F_l, \qquad (5)$$

where *c.i.t.* denotes the cubic interaction terms ($\propto \omega_{ch}\, a_{k'}\, a_{k''}\, a^*_{k'''}$, with $k'+k''-k'''=l$). For the explicit calculation of the coefficients $a_k$, see our Supplemental Material. The first term of Eq. (5) is linear in $a_l$ and becomes dominant far away from resonance and bifurcation. At resonance and close to bifurcation, for a single-frequency forcing $F_k$ a self-interaction term $|a_k|^2 a_k$ remains and the response $z \propto F^{1/3}$ emerges. In the presence of a second stimulus $F_{k+1}$, the response with respect to $F_k$ is suppressed (due to an interaction term $2\omega_{ch}a_k|a_{k+1}|^2$ that is linear in $a_k$), while at frequency $\omega_{k-1}$, the $2f_1-f_2$ CT is generated (via the interaction term $w_{ch}a_k^2 a^*_{k+1}$). Further cubic CT are generated at frequencies $\omega_l, l < k-1$. These are the strongest CT; for the following, we shall concentrate on them. Their amplitudes $a$ decrease according to $a(l) \propto \left(\frac{\omega_{ch}}{\omega_0}|a_1||a_2|\right)^{k-1-l}$. While the term $\propto (\omega_l - \omega_{ch})$ decreases with $l$ in steps of $\omega_0$, this is counter-balanced by the increasing number of contributing interaction terms. From this, a series of CT with exponentially decaying amplitudes emerges. The obtained decay exponents are corroborated by numerical integration of Eq. (4) for single Hopf elements. In the biological example as well as in the compound cochlea, the exponential decay is present as well, but with coherent consistently smaller exponents (cf. Fig. 1b). Isolated Hopf elements therefore describe the perceived pitch poorly, even at the Hopf bifurcation (see Supplemental Material).

*The role of the fluid* The 'up-validation' of the weaker CT-amplitudes (l.h.s decays in Fig. 1b), c)) could be based on two effects, both related to the presence of the cochlear fluid: 1) the feed-forward coupling of the Hopf amplifiers, or 2) the viscous damping acting on the cochlear fluid. Insight into their respective influence is obtained by running the cochlea without viscous damping. The corresponding experiment yields un-biological tuning curves (the part of the response associated with lower frequencies than $f_{ch}$ rises correctly, but the r.h.s part responsible for higher frequencies than $f_{ch}$ does not decay properly), but the extracted perceived pitch (its extraction see below) still follows the behavior predicted by Eq. (2) with $k' = k+1$, implying the fluid as the origin of the second pitch shift. In biology, CT of frequencies lower than stimulus frequencies propagate down the cochlea until the corresponding waves are amplified and stopped where their frequency matches the characteristic frequency $f_{ch}$. Therefore we first check that in our model CT amplification by the following sections reproduces the biological exponential scaling far from resonance, i.e. shows low-pass filtering with larger slopes of CT levels for frequencies above the stimulus frequencies. Up to discretization effects, the results obtained indeed fully coincide with the biological observations (Fig. 1c)). We now scrutinize the behavior of the strongest, the $2f_1-f_2$ CT, that is the direct product of the two interacting modes ($f_1$ and $f_2$). We compare the amplitudes evoked by a single pure tone $f_{ch}$ to a two-tone stimulation $f_1$, $f_2$ yielding a CT $2f_1 - f_2 = f_{ch}$, for the compound cochlea at the sixth cochlea section and a single Hopf oscillator where the fluid is not included in the model. The results (Fig. 2a), b), respectively) demonstrate that CT-amplitude varies nontrivially with stimulation level. A conventional quantification of the difference between the two inputs is the 'relative strength' of the $2f_1 - f_2$ tone. The relative strength is obtained by subtracting the primary level from the CT "equivalent level", defined as the stimulation level of a single pure tone at CT-frequency able to elicit a response of the same magnitude [12] (the primaries $f_1$ and $f_2$ are chosen at equal strength). The horizontal distances between the black and the green lines in Figs. 2a), b) illustrate this measure. Clearly, the compound system generates a qualitatively changed CT response: Upon feeding a general two-tone stimulus $2f_1 - f_2 = f_{ch}$[(6)] into the cochlea, the responses evoked by the primary frequencies $a_1$ and $a_2$ first grow linearly (1 dB/dB). Further down the cochlear duct, the cubic CT $a_{CT}$ grows at 3 dB/dB, whereas, towards the location of the respective characteristic frequencies, the primary responses enter the compressive regime (slope < 1 dB/dB). This observation translates directly into a strongly reduced growth of CT-response $a_{CT}$. Moreover, the response is further suppressed by the now substantial interaction terms $-2\omega_{ch}a_{CT}|a_1|^2$ and $-2\omega_{ch}a_{CT}|a_2|^2$ in the Hopf equation. Both effects force the CT to grow less than 1 dB/dB, with mostly a stronger compression than the rate of an equivalent pure tone at CF (black line to the left of Fig. 2 b)). The effect persists at more apical cochlear sites and fully complies with the biological measurements [22]. Whereas the relative

CT strength in isolated Hopf oscillators *increases* with input level and may only saturate/decrease at very high sound levels (Fig. 2 a)), in biology and in the Hopf cochlea, the relative strength *decreases* with input level (Fig. 2 b) or c)). For example, at a stimulation strength of -74 dB, no CT contribution is visible in the isolated case, in the compound cochlea the response is considerable instead (-40 dB).

*Pitch extraction* Our final step consists in the automated extraction of the perceived pitch from the model. In biology as well as in the cochlea model, two key factors determine the extracted pitch. 1) the stimulus amplitudes determine the CT decay exponents, 2) the choice of the cochlear pitch read-out place keeps track of the fluid's low-pass filtering and the overall-amplification of the lower CT. These features are not independent. Motivated by the *missing fundamental* paradigm and experiments by Smoorenburg [1], we concluded that the optimal read-out place should be given by the location of the lowest audible CT. Smoorenburg evaluated the lower limit of perception of the cubic CT to 40 dB SPL. The frequency of this CT then would indicate where along the cochlear duct the pitch should be extracted. The procedure (depicted in Fig. 3 a)) yields a read-out place that shifts monotonically with stimulation frequency (over ranges from around 700 to 1600 Hz, for our cochlea covering a frequency range from 440-14080 Hz). Within the limitations of the model's discretization, the agreement between model and biology is excellent: in Fig. 3 the full red line indicating the lowest audible CT (i.e., above −53 dB), is less than one section apart from the psychoacoustic data, and across all sections, the lowest perceptible CT is always the lowest CT above the psychoacoustic limit. The value of −53 dB is essentially chosen to obtain an optimal agreement with the psychophysical experiments (Fig. 3, a different choice of the hearing threshold would result in a parallel shift of the red curve). Identifying biophysical pitches with the psychoacoustically perceived pitch requires the faithfulness of the signal transduction from the cochlea via inner hair cells and the cochlea nerve neurons to the auditory cortex. A recent modeling study by us [13] demonstrated that the information available at the level of the cochlea is transmitted as accurately as one might imagine to the end of the auditory nerve. This is the final justification for comparing psychoacoustic pitch-shift experiments with the biophysical results. With the setting described in the last paragraph, Smoorenburg's corresponding psychoacoustic results (Fig. 4a)) and available pitch-shift related physiological data (Fig. 4b)) can be fully reproduced, without any further tuning. In the reproduction of Smoorenburg's psychoacoustical two-tone pitchshift experiments, the response signal to an input of the form $F_1 e^{2\pi i f_1 t} + F_2 e^{2\pi i (f_1+200)t}$ was read-out from the cochlea section determined as described (cf. Fig. 1a), red circles) and $f_p$ was computed from the dominant peaks of the signal's autocorrelation function. For both partials, a sound level of -74 dB was chosen, which corresponds to the partial levels of 40 dB SPL (see scale-correspondence in Fig. 1b), c)). Similarly, pitch correlate data from neurons of the cat cochlear nucleus [14] is reproduced. In this case, the biological measurement is from a single neuron having its own preferred frequency, which already sets the cochlea's read-out place.

*Conclusion* It has for a long time been an unresolved matter how pitch should be defined, where it is located and how it should be extracted [2, 23, 24]. Here we provided for the first time a reproduction and a quantitative explanation of the second pitch-shift, based on a simple physical model of cochlea. From the confirmed close correspondence between biology and model, we revealed the key roles played by CT and by the cochlear fluid. Upon reproducing Smoorenburg's psychoacoustic pitch-shift, the read-out place shifted monotonically with the primary frequencies; in the case of the pitch correlate data from the cat cochlear nucleus [14] (Fig. 4b)), the read-out place was set by the neuron's characteristic firing frequency. Both comparisons request the faithfulness of the inner hair cell-auditory nerve signal transduction, a credo in the field that we have recently corroborated by modeling evidence (cf. [13]). The presented work establishes the second pitch-shift as a purely biophysical phenomenon. Working from the sensor inwards, a similar understanding of the next afferent stations of the auditory pathway may provide fundamental insights into brain function. One may ask oneself what part in pitch perception will finally be left to the brain? In our study, all Hopf parameters were held at their fixed predefined values. However, cascades of feedback loops from the cortex to the cochlea provide efferent (generally: inhibitory) input to the cochlea, which will modify unspecific Hopf parameter patterns. In mixtures of sounds, specific parameter patterns allow focusing on individual auditory objects that have their own characteristic pitch [25]. The choice of auditory object then, would be left to the brain.

[1] G. F. Smoorenburg, J. Acoust. Soc. Am. **48**, 924 (1970).

[2] A. Cheveign´e, in *Pitch*, edited by C. Plack, R. Fay, A. Oxenham, and A. Popper (Springer, New York, 2005), pp. 169–233.

[3] A. Seebeck, Ann. Phys. Chem. **53**, 417 (1841).

[4]  J. F. Schouten, Philips Technisch Tijdschr. **5**, 298 (1940).

[5]  E. de Boer, Nature **178**, 535 (1956).

[6]  J. F. Schouten, R. J. Ritsma, and B. L. Cardozo, J. Acoust. Soc. Am. **34**, 1418 (1962).

[7]  D. R. Chialvo, O. Calvo, D. L. Gonzalez, O. Piro, and G. V. Savino, Phys. Rev. E **65**, 050902 (2002).

[8]  J. Licklider, Cell. Mol. Life. Sci. **7**, 128 (1951).

[9]  P. A. Cariani and B. Delgutte, J. Neurophysiol. **76**, 1698 (1996).

[10] J. Goldstein and N. Kiang, Proc. IEEE **56**, 981 (1968).

[11] J. L. Goldstein, J. Acoust. Soc. Am. **41**, 676 (1967).

[12] L. Robles, M. A. Ruggero, and N. C. Rich, J. Neurophysiol. **77**, 2385 (1997).

[13] S. Martignoli, F. Gomez, and R. Stoop, Scientific reports **3**, 2676 (2013).

[14] W. S. Rhode, J. Acoust. Soc. Am. **97**, 2414 (1995).

[15] V. M. Eguiluz, M. Ospeck, Y. Choe, A. J. Hudspeth, and M. O. Magnasco, Phys. Rev. Lett. **84**, 5232 (2000).

[16] S. Camalet, T. Duke, F. Jülicher, and J. Prost, Proc. Natl. Acad. Sci. U.S.A. **97**, 3183 (2000).

[17] A. Kern and R. Stoop, Phys. Rev. Lett. **91**, 128101 (2003).

[18] R. Stoop and A. Kern, Phys. Rev. Lett. **93**, 268103 (2004).

[19] S. Martignoli, J.-J. van der Vyver, A. Kern, Y. Uwate, and R. Stoop, Appl. Phys. Lett. **91**, 064108 (2007).

[20] S. Martignoli and R. Stoop, Phys. Rev. Lett. **105**, 048101 (2010).

[21] E. Zwicker, Acustica **5**, 67 (1955).

[22] N. P. Cooper and W. S. Rhode, J. Neurophysiol. **78**, 261 (1997).

[23] R. Plomp, J. Acoust. Soc. Am. **41**, 1526 (1967).

[24] R. J. Ritsma, J. Acoust. Soc. Am. **42**, 191 (1967).

[25] F. Gomez, V. Saase, N. Buchheim, and R. Stoop, submitted (2013).

**Author contributions:**

The work was achieved by the authors as a team-work.

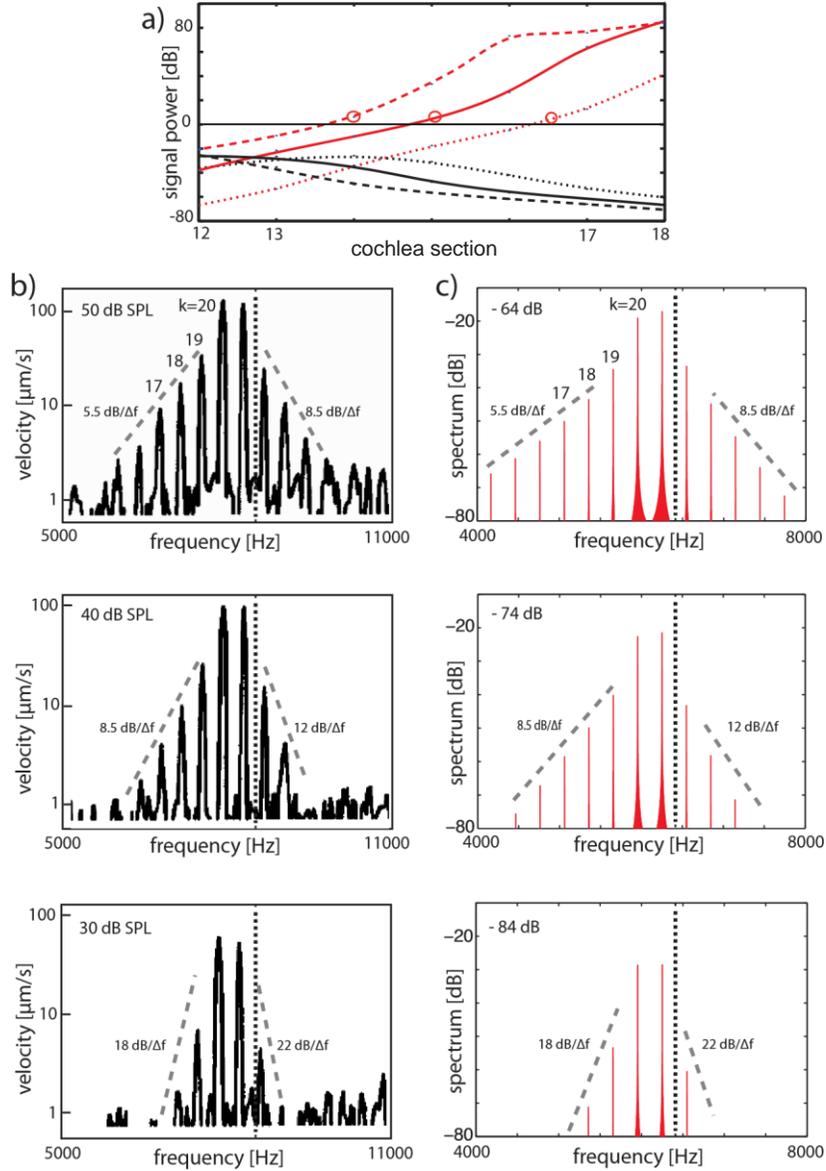

FIG. 1: a) CT-power dominance along the cochlea (section 12 = 1893 Hz to section 18 = 634 Hz). Two-tone stimulation with $f_1, f_2$ = 1200, 1400 Hz (dotted); 1600, 1800 Hz (solid); 2000, 2200 Hz (dashed) at -74 dB (40 dB SPL) sound level. Red: total cochlear CT-power lower than $f_1$ divided by the sum of the power at $f_1, f_2$, measured at the respective sections. Black: total cochlear signal power, measured at the respective sections. Red circles indicate the conditions relevant for the pitch-read out. b) Basilar membrane response spectrograms for two-tone stimulation of amplitudes 30, 40, 50 dB SPL (frequencies $f_2/f_1$ = 1.05 and $2f_2 - f_1 = f_{ch}$) in the cochlear apex, biological data [12] ($f_{ch}$ = 7500

Hz), c) Hopf-cochlea model, 6th section ($f_{ch}$ = 5656 Hz). Relevant forcing and lower CT frequencies are left to the dotted lines; dashed lines: exponential amplitude scaling ($\Delta f = f_2 - f_1$).

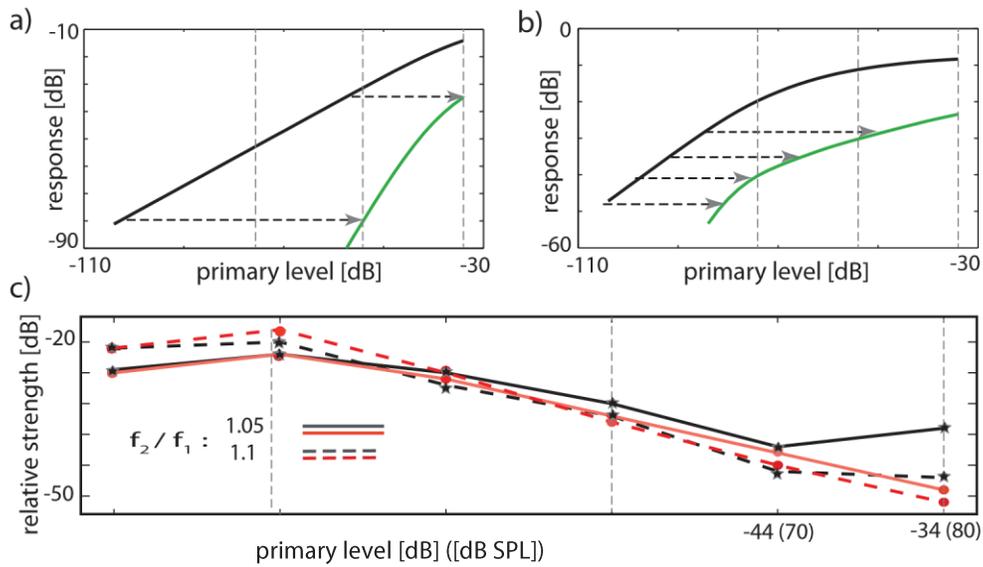

FIG. 2: Response at characteristic frequency $f_{ch}$ = 5656 Hz, when stimulated by a signal $f_{ch}$ (black) and by two pure tones $f_1 f_2$ of same strength (green) generating a CT $2f_1 - f_2 = f_{ch}$ of same output strength, a) single Hopf amplifier (no signal propagation), b) cochlea section 6. Arrows indicate relative strength. c) Relative strength of the $2f_1$–$f_2$ CT as a function of primary level (green curve) for two $f_2/f_1$-frequency ratios. Red: cochlea section 6, black: biological data [12] ($f_{ch}$ = 9000 Hz): The compound model reproduces the biological data very well; isolated Hopf elements won't.

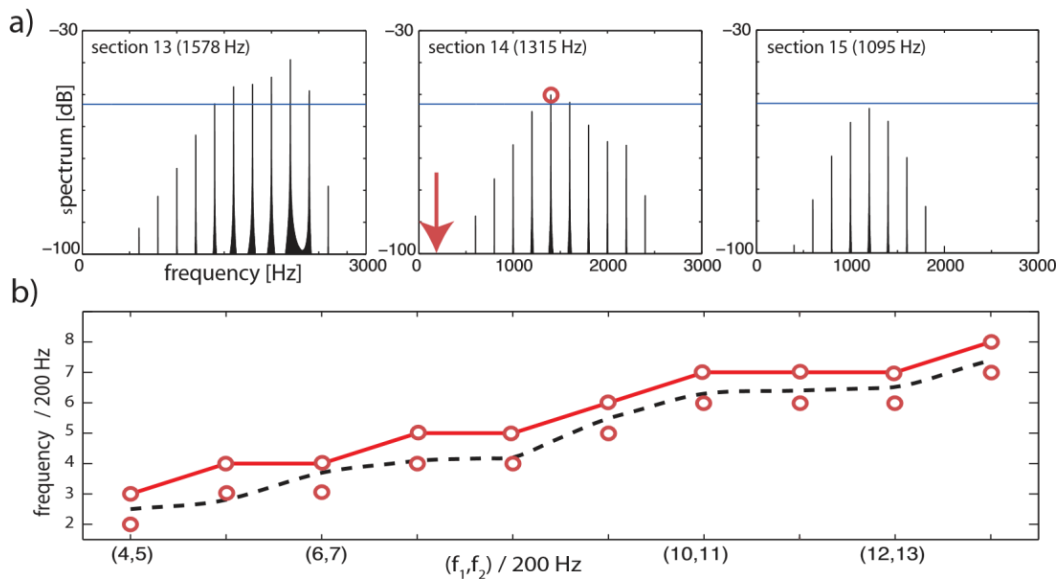

FIG. 3: a) Spectra for two-tone stimulation (-74 dB, $f_1$ = 2200 Hz, $f_2$ = 2400 Hz) at three cochlea sections. The lowest audible CT (hearing threshold: -53 dB, blue line) is the response at 1400 Hz (section 14, circled). The perceived pitch is the *residue pitch* (red arrow) associated with the spectrum at this location. b) Dashed black line: psychoacoustical lower hearing frequency limit of CT[1]. Full red line: lowest CT above the implemented amplitude threshold, single circles: highest CT below the limit. Across the whole frequency range, the three characteristics differ less than the width of a section.

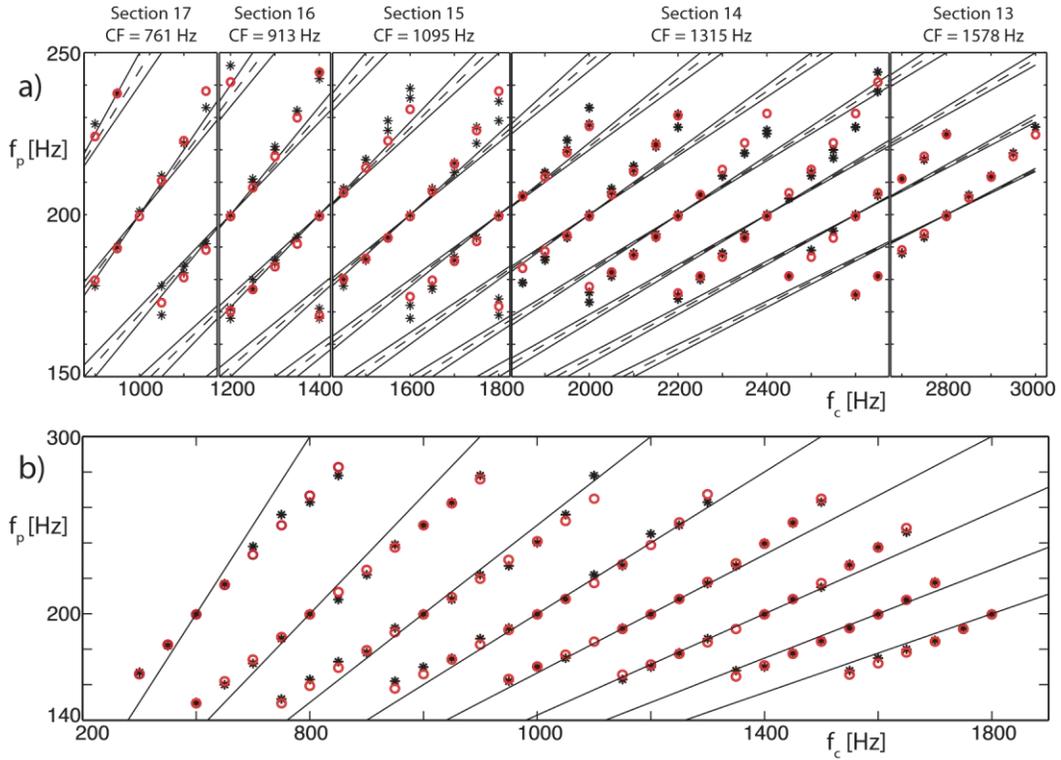

FIG. 4: a) Smoorenburg's pitch-shift experiment. Two-frequency stimulation $f_2 = f_1$ + 200 Hz. Black stars: psychoacoustic data [1] (partial sound levels 40 dB SPL, two subjects), red dots: Hopf cochlea (sections as indicated, partial tones -74 dB each). Black lines: false predictions by Eq. (2) for $k' = k$, $k' = k + 1/2$ (dashed) and $k' = k$+ 1, respectively. b) Cat ventral cochlear nucleus. Three-frequency stimulation ($f_c - f_{mod}$), $f_c$ ,($f_c + f_{mod}$) ($f_{mod}$ = 200 Hz). Black stars: inverse of most frequent ISI, On-L-cell response [14] (CF= 1100 Hz, 50 dB SPL, red dots: pitch from cochlea section 15 (CF= 1095 Hz, -64 dB).

12